\documentclass[12pt,draft]{revtex4}

\usepackage{amsmath}
\usepackage{amsfonts}
\usepackage{latexsym}
\usepackage{amssymb} 
\usepackage{verbatim}
\usepackage{graphicx}

\newtheorem{conjecture}{Conjecture}[section]

\begin{document}

\markboth{K.~Nakamura}
{Alternative construction of gauge-invariant variables on general background spacetime}

%
%

\title{
  Alternative construction of gauge-invariant variables\\
  for linear metric perturbation\\
  on general background spacetime
}

\author{
  Kouji Nakamura\footnote{e-mail:kouji.nakamura@nao.ac.jp}
}

\address{
  Optical and Infrared Astronomy Division,
  National Astronomical Observatory of Japan,\\
  Osawa 2-21-1, Mitaka 181-8588, Japan
}


\date{March 16, 2011}

\begin{abstract}
  Construction of the gauge-invariant variables for the linear
  metric perturbation, which was proposed in the paper
  [K.~Nakamura, arXiv:1101.1147], is discussed through an
  alternative approach.
  Our starting point of the construction of the gauge-invariant
  variables is an non-trivial non-local decomposition of the
  linear metric perturbation.
  Assuming the existence of some Green functions, we reproduce
  results in the above paper.
  This supports the consistency of the result and implies that
  one can develop the general-relativistic higher-order
  gauge-invariant perturbation theory on general background
  spacetime.
\end{abstract}


\maketitle

\section{Introduction}
\label{sec:intro}


General relativity is a theory in which the construction of
exact solutions is not so easy.
In this situation, perturbation theories are powerful techniques
and the developments of perturbation theories lead physically
fruitful results and interpretations of natural phenomena. 
For this reason, general relativistic {\it linear} perturbation
theory has been widely used in many area\cite{Bardeen-1980}.
Further, the investigation of {\it higher-order}
general-relativistic perturbations is also necessary due to the
precise observations\cite{Non-Gaussianity-observation} in
cosmology\cite{Tomita-1967-Non-Gaussianity,kouchan-cosmo-second},
and to prepare more precise wave form of gravitational
wave\cite{Gleiser-Nicasio} for the gravitational wave
detection.


As well-known, general relativity is based on the concept of
general covariance.
Due to this general covariance, the ``gauge degree of freedom'', 
which is an unphysical degree of freedom, arises in
general-relativistic perturbations.
To obtain physical results, we have to fix this gauge degrees of
freedom or to extract some invariant quantities of
perturbations.
This situation becomes more complicated in higher-order
perturbations. 
Since there are so-called {\it gauge-invariant} linear
perturbation theories in some background spacetimes, it is
worthwhile to investigate higher-order gauge-invariant
perturbation theory from a general point of view to avoid this
gauge issues.


According to these motivation, the general framework of
higher-order general-relativistic gauge-invariant perturbation
theory has been discussed in some papers by the present
author\cite{kouchan-gauge-inv}.
In this general framework, we consider the perturbative
expansion of the physical metric 
$\bar{g}_{ab}=g_{ab}+\lambda h_{ab}+O(\lambda^{2})$, where
$\lambda$ is an infinitesimal parameter for the perturbation
theory.
$h_{ab}$ is the linear-order metric perturbation.
Further, in our general framework, we assumed the following
conjecture:


\begin{conjecture}
  \label{conjecture:decomposition-conjecture}
  If there is a tensor field $h_{ab}$ of the second rank, whose
  gauge transformation rule is
  \begin{eqnarray}
    {}_{{\cal Y}}\!h_{ab}
    -
    {}_{{\cal X}}\!h_{ab}
    =
    {\pounds}_{\xi}g_{ab},
    \label{eq:linear-metric-gauge-trans}
  \end{eqnarray}
  then there exist a tensor field ${\cal H}_{ab}$ and a vector
  field $X^{a}$ such that $h_{ab}$ is decomposed as 
  \begin{eqnarray}
    h_{ab} =: {\cal H}_{ab} + {\pounds}_{X}g_{ab},
    \label{eq:linear-metric-decomp}
  \end{eqnarray}
  where ${\cal H}_{ab}$ and $X^{a}$ are transformed as
  \begin{equation}
    {}_{{\cal Y}}\!{\cal H}_{ab} - {}_{{\cal X}}\!{\cal H}_{ab} =  0, 
    \quad
    {}_{\quad{\cal Y}}\!X^{a} - {}_{{\cal X}}\!X^{a} = \xi^{a}.
    \label{eq:linear-metric-decomp-gauge-trans}
  \end{equation}
  Here, ${\cal Y}$ and ${\cal X}$ denote different gauge choices. 
\end{conjecture}


In the case of cosmological perturbations, we confirmed that
this conjecture is almost true, and then, we developed the
second-order gauge-invariant cosmological
perturbations\cite{kouchan-cosmo-second}.
Since our general framework of higher-order perturbations does
not depend on details of the background metric $g_{ab}$, we will
be able to develop general-relativistic higher-order
gauge-invariant perturbation theory if Conjecture 
\ref{conjecture:decomposition-conjecture} is
true\cite{kouchan-gauge-inv}.
Although we recently propose a scenario of the proof of
Conjecture \ref{conjecture:decomposition-conjecture}, in this
article, we give an alternative explanation of this proof.
This is the main purpose of this article.


\section{Construction of gauge-invariant variables}
\label{sec:construction}


Now, we give a scenario of a proof of Conjecture
\ref{conjecture:decomposition-conjecture} on general background 
spacetimes.
Here, we assume that the background spacetimes considered in
this paper admit ADM decomposition\cite{Wald-book}.
Therefore, the background spacetime ${\cal M}_{0}$ considered
here is $n+1$-dimensional spacetime which is foliated by
spacelike hypersurfaces $\Sigma(t)$ ($\dim\Sigma = n$).
Each $\Sigma$ may have its boundary $\partial\Sigma$.
The background metric $g_{ab}$ is given by 
\begin{eqnarray}
  \label{eq:gdb-decomp-dd-minus-main}
  g_{ab} &=& - \alpha^{2} (dt)_{a} (dt)_{b}
  + q_{ij}
  (dx^{i} + \beta^{i}dt)_{a}
  (dx^{j} + \beta^{j}dt)_{b},
\end{eqnarray}
where $\alpha$ is the lapse function, $\beta^{i}$ is the
shift vector, and $q_{ab}=q_{ij}(dx^{i})_{a}(dx^{i})_{b}$ is the
metric on $\Sigma(t)$.


To consider the decomposition (\ref{eq:linear-metric-decomp}) of
$h_{ab}$, we, first, consider the components of the metric
$h_{ab}$ as
\begin{eqnarray}
  h_{ab}
  =
  h_{tt} (dt)_{a}(dt)_{b}
  + 2 h_{ti} (dt)_{(a}(dx^{i})_{b)}
  + h_{ij} (dx^{i})_{a}(dx^{j})_{b}.
  \label{eq:K.Nakamura-2010-note-1-2}
\end{eqnarray}
The gauge transformation rule
(\ref{eq:linear-metric-gauge-trans}) gives the transformation
rules for the components $\{h_{tt},h_{ti},h_{ij}\}$, which are
given by
\begin{eqnarray}
  {}_{{\cal Y}}h_{tt}
  -
  {}_{{\cal X}}h_{tt}
  &=&
    2 \partial_{t}\xi_{t}
  - \frac{2}{\alpha}\left(
    \partial_{t}\alpha 
    + \beta^{i}D_{i}\alpha 
    - \beta^{j}\beta^{i}K_{ij}
  \right) \xi_{t}
  \nonumber\\
  &&
  - \frac{2}{\alpha} \left(\frac{}{}
    \beta^{i}\beta^{k}\beta^{j} K_{kj}
    - \beta^{i} \partial_{t}\alpha
    + \alpha q^{ij} \partial_{t}\beta_{j}
  \right.
  \nonumber\\
  && \quad\quad\quad
  \left.
    + \alpha^{2} D^{i}\alpha 
    - \alpha \beta^{k} D^{i} \beta_{k}
    - \beta^{i} \beta^{j} D_{j}\alpha 
    \frac{}{}
  \right)\xi_{i}
  \label{eq:K.Nakamura-2010-note-3-2}
  , \\
  {}_{{\cal Y}}h_{ti}
  -
  {}_{{\cal X}}h_{ti}
  &=&
  \partial_{t}\xi_{i}
  + D_{i}\xi_{t}
  - \frac{2}{\alpha} \left(
    D_{i}\alpha 
    - \beta^{j}K_{ij}
  \right) \xi_{t}
  \nonumber\\
  &&
  - \frac{2}{\alpha} \left(
    - \alpha^{2} K^{j}_{\;\;i}
    + \beta^{j}\beta^{k} K_{ki}
    - \beta^{j} D_{i}\alpha 
    + \alpha D_{i}\beta^{j}
  \right) \xi_{j}
  \label{eq:K.Nakamura-2010-note-4-2}
  , \\
  {}_{{\cal Y}}h_{ij}
  -
  {}_{{\cal X}}h_{ij}
  &=&
  2 D_{(i}\xi_{j)}
  + \frac{2}{\alpha} K_{ij} \xi_{t}
  - \frac{2}{\alpha} \beta^{k} K_{ij} \xi_{k}
  ,
  \label{eq:K.Nakamura-2010-note-5-2}
\end{eqnarray}
where $K_{ij}$ is the extrinsic curvature:
\begin{eqnarray}
  \label{eq:K.Nakamura-2010-2-generic-alternative-logic-2.21}
  K_{ij} = - \frac{1}{2\alpha}
  \left[\frac{\partial}{\partial t} q_{ij} -
  D_{i}\beta_{j} - D_{j}\beta_{i}\right],
\end{eqnarray}
and $D_{i}$ is the covariant derivative associated with the
spatial metric $q_{ij}$.
Inspecting these gauge transformation rules and assuming the
existence of the Green function of the derivative operator
$\Delta:=D^{i}D_{i}$, we consider the decompositions of the
components $h_{ti}$ and $h_{ij}$ as follows:
\begin{eqnarray}
  h_{ti}
  &=:&
  D_{i}h_{(VL)} + h_{(V)i}
  - \frac{2}{\alpha} \left(
    D_{i}\alpha 
    - \beta^{j}K_{ij}
  \right) h_{(VL)}
  \nonumber\\
  &&
  - \frac{2}{\alpha} \left(
    D_{i}\alpha 
    - \beta^{j}K_{ij}
  \right)
  \Delta^{-1}
  \left[
    \partial_{t}D^{k}h_{(TV)k}
    - 2 D_{l}\left\{
      h_{(TV)k} \left( \alpha K^{kl} - D^{(k}\beta^{l)} \right)
    \right\}
  \right.
  \nonumber\\
  && \quad\quad\quad\quad\quad\quad\quad\quad\quad\quad\quad
  \left.
    +   h_{(TV)k} D^{k}\left( \alpha K - D^{l}\beta_{l} \right)
    \frac{}{}
  \right]
  \nonumber\\
  &&
  - \frac{2}{\alpha} \left(
    - \alpha^{2} K^{j}_{\;\;i}
    + \beta^{j}\beta^{k} K_{ki}
    - \beta^{j} D_{i}\alpha 
    + \alpha D_{i}\beta^{j}
  \right) h_{(TV)j}
  \label{eq:K.Nakamura-2010-2-generic-alternative-logic-2.27}
  , \\
  h_{ij}
  &=:&
  \frac{1}{n} q_{ij} h_{(L)} + h_{(T)ij}
  + \frac{2}{\alpha} K_{ij} h_{(VL)}
  - \frac{2}{\alpha} \beta^{k} K_{ij} h_{(TV)k}
  \nonumber\\
  &&
  - \frac{2}{\alpha} K_{ij} \Delta^{-1}
  \left[\frac{}{}
    \partial_{t}D^{k}h_{(TV)k}
    - 2 D_{l}\left\{
      h_{(TV)k} \left( \alpha K^{kl} - D^{(k}\beta^{l)} \right)
    \right\}
  \right.
  \nonumber\\
  && \quad\quad\quad\quad\quad\quad
  \left.
    +   h_{(TV)k} D^{k}\left( \alpha K - D^{l}\beta_{l} \right)
    \frac{}{}
  \right]
  ,
  \label{eq:K.Nakamura-2010-2-generic-alternative-logic-2.29}
  \\
  h_{(T)ij} &=:& D_{i}h_{(TV)j} + D_{j}h_{(TV)i}
  - \frac{2}{n} q_{ij} D^{l}h_{(TV)l}
  + h_{(TT)ij},
  \label{eq:K.Nakamura-2010-2-generic-alternative-logic-2.30}
  \\
  && D^{i}h_{(V)i} = 0, \quad
  q^{ij}h_{(TT)ij} = 0, \quad
  D^{i}h_{(TT)ij} = 0.
  \label{eq:K.Nakamura-2010-2-generic-alternative-logic-2.28}
\end{eqnarray}


From the gauge-transformation rule
(\ref{eq:K.Nakamura-2010-note-3-2}) and the definitions
(\ref{eq:K.Nakamura-2010-2-generic-alternative-logic-2.27}) and
(\ref{eq:K.Nakamura-2010-2-generic-alternative-logic-2.28}) of
the variable $h_{(LV)}$ and $h_{(V)i}$, we can derive the
gauge-transformation rules for these variables by assuming the
existence of the Green function ${\cal F}^{-1}$ of the elliptic
derivative operator
\begin{eqnarray}
  \label{eq:K.Nakamura-2010-2-generic-alternative-logic-2.36}
  {\cal F}
  := 
  \Delta
  - \frac{2}{\alpha} \left(
    D_{i}\alpha 
    - \beta^{j}K_{ij}
  \right)D^{i}
  - 2 D^{i}\left\{
    \frac{1}{\alpha} \left(
      D_{i}\alpha 
      - \beta^{j}K_{ij}
    \right)
  \right\}.
\end{eqnarray}
However, in these gauge-transformation rules, the variable
$h_{(TV)i}$ defined in
Eqs.~(\ref{eq:K.Nakamura-2010-2-generic-alternative-logic-2.29})--(\ref{eq:K.Nakamura-2010-2-generic-alternative-logic-2.28})
is also included.
This means that the gauge-transformation rules for variables
$h_{(LV)}$ and $h_{(V)i}$ are specified if we have the
gauge-transformation rule for $h_{(TV)i}$. 
From the trace part of the gauge-transformation rule
(\ref{eq:K.Nakamura-2010-note-5-2}), we can derive the
gauge-transformation rule for the variable $h_{(L)}$, which is
defined by
Eqs.~(\ref{eq:K.Nakamura-2010-2-generic-alternative-logic-2.29})--(\ref{eq:K.Nakamura-2010-2-generic-alternative-logic-2.28}).
This gauge-transformation rule also includes the variables
$h_{(LV)}$ and $h_{(TV)i}$.
Since the gauge-transformation rule for the variable $h_{(LV)}$
is determined if the gauge-transformation rule for the variable
$h_{(TV)i}$ is specified, the gauge-transformation rule for
$h_{(L)}$ is also specified if the gauge-transformation rule for
$h_{(TV)i}$ is given.
Thus, the gauge-transformation rules for the variables
$h_{(LV)}$, $h_{(V)i}$, and $h_{(L)}$ are specified if the
gauge-transformation for the variable $h_{(TV)i}$ is specified.


Taking the divergence of the traceless part of the
gauge-transformation rule (\ref{eq:K.Nakamura-2010-note-5-2})
and using
Eqs.~(\ref{eq:K.Nakamura-2010-2-generic-alternative-logic-2.29})--(\ref{eq:K.Nakamura-2010-2-generic-alternative-logic-2.28})
and the gauge-trasformation rule for the variable $h_{(LV)}$
obtained above, we reach to a single equation for a unknown
single vector field 
$A_{i}:={}_{{\cal Y}}h_{(TV)i}-{}_{{\cal X}}h_{(TV)i}-\xi_{i}$:
\begin{eqnarray}
  {\cal D}^{jl} A_{l}
  &=& D_{i}\left[
    \frac{2}{\alpha} \widetilde{K}^{ij}
    \left\{\frac{}{}
      {\cal F}^{-1}\left[\frac{}{}
        \partial_{t}D^{k}A_{k}
        - 2 D_{l}\left\{
          \frac{}{}
          A_{k}
          \left( \alpha K^{kl} - D^{(k}\beta^{l)} \right)
          +
          \frac{1}{\alpha} L^{ml} A_{m}
        \right\}
      \right.
    \right.
  \right.
  \nonumber\\
  && \quad\quad\quad\quad\quad\quad\quad\quad
  \left.
    \left.
      \left.
        + A_{k} D^{k}\left( \alpha K - D^{l}\beta_{l} \right)
        \frac{}{}
      \right]
    \right.
  \right.
  \nonumber\\
  && \quad\quad\quad\quad\quad
  \left.
    \left.
      - \beta^{k}A_{k}
      \frac{}{}
    \right\}
  \right]
  ,
  \label{eq:Ai-master-eq}
\end{eqnarray}
where the derivative operator ${\cal D}^{ij}$ and the
tensors $\tilde{K}^{ij}$ and $L^{ij}$ are defined by 
\begin{eqnarray}
  {\cal D}^{ij}
  &:=&
  q^{ij}\Delta + \left(1 - \frac{2}{n}\right) D^{i}D^{j} +
  R^{ij},
  \quad
  \widetilde{K}^{ij} 
  :=
  K^{ij} - \frac{1}{n} q^{ij} K
  ,
  \quad
  K := q^{ij}K_{ij}, 
  \nonumber\\
  L^{ij}
  &:=&
  - \alpha^{2} K^{ij}
  + \beta^{i}\beta_{k} K^{kj}
  - \beta^{i} D^{j}\alpha
  + \alpha D^{j}\beta^{i}
  .
  \label{eq:J.W.York.Jr-1973-1-2-kouchan-3-main}
\end{eqnarray}
As a trivial solution to Eq.~(\ref{eq:Ai-master-eq}), we have
$A_{i}=0$, i.e., 
\begin{eqnarray}
  {}_{{\cal Y}}h_{(TV)i} - {}_{{\cal X}}h_{(TV)i} = \xi_{i}
  \label{eq:K.Nakamura-2010-2-generic-alternative-logic-2.56}
\end{eqnarray}
Thus, we have obtained the gauge-transformation rule for the
variable $h_{(TV)i}$.


Using the gauge-transformation rule
(\ref{eq:K.Nakamura-2010-2-generic-alternative-logic-2.56}) for
the variable $h_{(TV)i}$, the gauge-transformation rules for
variables $h_{tt}$, $h_{(VL)}$, $h_{(V)i}$, $h_{(L)}$, and $h_{(TT)ij}$
are given by
\begin{eqnarray}
  {}_{{\cal Y}}h_{(VL)}
  -
  {}_{{\cal X}}h_{(VL)}
  &=&
  \xi_{t}
  + \Delta^{-1} \left[
    \partial_{t}D^{k}\xi_{k}
    - 2 D_{l}\left\{ \xi_{k} \left( \alpha K^{kl} - D^{(k}\beta^{l)} \right) \right\}
  \right.
  \nonumber\\
  && \quad\quad\quad\quad\quad
  \left.
    + \xi_{k} D^{k}\left( \alpha K - D^{l}\beta_{l} \right)
    \frac{}{}
  \right]
  \label{eq:K.Nakamura-2010-2-generic-alternative-logic-2.62}
  , \\
  {}_{{\cal Y}}h_{(V)i}
  -
  {}_{{\cal X}}h_{(V)i}
  &=&
  \partial_{t}\xi_{i}
  - D_{i}\Delta^{-1}\left[
    \partial_{t}D^{k}\xi_{k}
    - 2 D_{l}\left\{
      \xi_{k} \left( \alpha K^{kl} - D^{(k}\beta^{l)} \right)
    \right\}
  \right.
  \nonumber\\
  && \quad\quad\quad\quad\quad\quad\quad
  \left.
    + \xi_{k} D^{k}\left( \alpha K - D^{l}\beta_{l} \right)
    \frac{}{}
  \right]
  ,
  \label{eq:K.Nakamura-2010-2-generic-alternative-logic-2.63}
  \\
  {}_{{\cal Y}}h_{(L)}
  -
  {}_{{\cal X}}h_{(L)}
  &=& 
  2 D^{i}\xi_{i}
  ,
  \label{eq:K.Nakamura-2010-2-generic-alternative-logic-2.64}
  \\
  {}_{{\cal Y}}h_{(TT)ij}
  -
  {}_{{\cal X}}h_{(TT)ij}
  &=&
  0.
  \label{eq:K.Nakamura-2010-2-generic-alternative-logic-2.66}
\end{eqnarray}


Inspecting the gauge-transformation rules
(\ref{eq:K.Nakamura-2010-2-generic-alternative-logic-2.56}) and
(\ref{eq:K.Nakamura-2010-2-generic-alternative-logic-2.62}), we
first construct the variables $X_{t}$ and $X_{i}$ which satisfy
the properties ${}_{{\cal Y}}X_{t}-{}_{{\cal X}}X_{t}=\xi_{t}$, 
${}_{{\cal Y}}X_{i}-{}_{{\cal X}}X_{i}=\xi_{i}$, respectively. 
We can easily find these variables as follows:
\begin{eqnarray}
  X_{t}
  &:=&
  h_{(VL)}
  - \Delta^{-1} \left[ \frac{}{}
    \partial_{t}D^{k}h_{(TV)k}
    - 2 D_{l}\left\{
      h_{(TV)k} \left( \alpha K^{kl} - D^{(k}\beta^{l)} \right)
    \right\}
  \right.
  \nonumber\\
  && \quad\quad\quad\quad\quad\quad\quad
  \left.
    + h_{(TV)k} D^{k}\left( \alpha K - D^{l}\beta_{l} \right)
    \frac{}{}
  \right]
  ,
  \label{eq:K.Nakamura-2010-2-generic-alternative-logic-2.70}
  \\
  X_{i} &:=& h_{(TV)i}.
  \label{eq:K.Nakamura-2010-2-generic-alternative-logic-2.69}
\end{eqnarray}
These $X_{t}$ and $X_{i}$ satisfy the desired
gauge-transformation rules, respectively.


Through these $X_{t}$ and $X_{i}$ and the gauge-transformation
rules (\ref{eq:K.Nakamura-2010-note-3-2}),
(\ref{eq:K.Nakamura-2010-2-generic-alternative-logic-2.63})--(\ref{eq:K.Nakamura-2010-2-generic-alternative-logic-2.66}),
we can easily define gauge-invariant variables $\Phi$, $\Psi$,
$\nu_{i}$, and $\chi_{ij}$ by
\begin{eqnarray}
  - 2 \Phi
  &:=&
  h_{tt}
  - 2 \partial_{t}X_{t}
  + \frac{2}{\alpha}\left(
    \partial_{t}\alpha 
    + \beta^{i}D_{i}\alpha 
    - \beta^{j}\beta^{i}K_{ij}
  \right) X_{t}
  \nonumber\\
  &&
  + \frac{2}{\alpha} \left( \frac{}{}
    \beta^{i}\beta^{k}\beta^{j} K_{kj}
    - \beta^{i} \partial_{t}\alpha
    + \alpha q^{ij} \partial_{t}\beta_{j}
    + \alpha^{2} D^{i}\alpha 
  \right.
  \nonumber\\
  && \quad\quad\quad
  \left.
    - \alpha \beta^{k} D^{i} \beta_{k}
    - \beta^{i} \beta^{j} D_{j}\alpha 
    \frac{}{}
  \right) X_{i}
  ,
  \label{eq:K.Nakamura-2010-2-generic-alternative-logic-2.81}
  \\
  - 2 n \Psi
  &:=&
  h_{(L)} - 2 D^{i}X_{i},
  \label{eq:K.Nakamura-2010-2-generic-alternative-logic-2.82}
  \\
  \nu_{i}
  &:=&
  h_{(V)i}
  - \partial_{t}X_{i}
  \nonumber\\
  &&
  + D_{i}\Delta^{-1}\left[\frac{}{}
    \partial_{t}D^{k}X_{k}
    - 2 D_{l}\left\{
      X_{k} \left( \alpha K^{kl} - D^{(k}\beta^{l)} \right)
    \right\}
  \right.
  \nonumber\\
  && \quad\quad\quad\quad\quad
  \left.
    + X_{k} D^{k}\left( \alpha K - D^{l}\beta_{l} \right)
    \frac{}{}
  \right]
  ,
  \label{eq:K.Nakamura-2010-2-generic-alternative-logic-2.83}
  \\
  \chi_{ij} &:=& h_{(TT)ij}.
  \label{eq:K.Nakamura-2010-2-generic-alternative-logic-2.84}
\end{eqnarray}
Actually, we can easily confirm that these variables are gauge
invariant. 
We also note that the variable $\nu_{i}$ and $\chi_{ij}$ satisfy
the properties $D^{i}\nu_{i}=0$, $q^{ij}\chi_{ij}=0$,
$\chi_{[ij]}=0$, and $D^{i}\chi_{ij}=0$.


Through the original components $h_{tt}$, $h_{ti}$, and
$h_{ij}$ in terms of the variables $\Phi$, $\Psi$, $\nu_{i}$,
$\chi_{ij}$, $X_{t}$, and $X_{i}$, the first-order metric
perturbation $h_{ab}$ are given in the form as
Eq.~(\ref{eq:linear-metric-decomp}) by choosing 
\begin{eqnarray}
  {\cal H}_{ab} 
  &=&
  - 2 \Phi (dt)_{a}(dt)_{b}
  + 2 \nu_{i} (dt)_{(a}(dx^{i})_{b)}
  + \left(
    - 2 q_{ij} \Psi + \chi_{ij}
  \right) (dx^{i})_{a}(dx^{j})_{b}
  \label{eq:K.Nakamura-2010-2-generic-alternative-logic-2.96}
  , \\
  X_{a} &=& X_{t} (dt)_{a} + X_{i}(dx^{i})_{a}
  \label{eq:K.Nakamura-2010-2-generic-alternative-logic-2.98}
  .
\end{eqnarray}
Thus, we have confirmed Conjecture
\ref{conjecture:decomposition-conjecture}.


\section{Summary}
\label{sec:discussions}


We have shown an alternative scenario of a proof of Conjecture
\ref{conjecture:decomposition-conjecture}. 
In our previous paper\cite{kouchan-decomp-previous}, we
first assumed the existence of the gauge-variant variables
$X_{t}$ and $X_{i}$ and confirmed this existence through the
explicit construction of these variables.
Logically speaking, this is a no-trivial logic and we have
explicitly constructed gauge-invariant variables $\Phi$, $\Psi$,
$\nu_{i}$, and $\chi_{ij}$.
In this previous derivation, we assume the existence of the
Green functions of the Laplacian $\Delta$ and the elliptic
derivative operator ${\cal D}_{ij}$ in
Eqs.~(\ref{eq:J.W.York.Jr-1973-1-2-kouchan-3-main}). 
Therefore, special modes which belong to the kernels of these
two derivative operators are excluded in our consideration.
To include these modes, different treatments will be necessary.
We called this problem as {\it zero-mode problem}.
In this paper, we gave an alternative construction of the
variable $X_{t}$ and $X_{i}$ to support our previous
result\cite{kouchan-decomp-previous}.
In this sense, we may say that the results in our previous
paper\cite{kouchan-decomp-previous} are correct.


However, in the approach in this article, we assumed the
existence of the Green function of the elliptic derivative
operator ${\cal F}$, which defined by
Eq.~(\ref{eq:K.Nakamura-2010-2-generic-alternative-logic-2.36}),
instead of ${\cal D}_{ij}$.
Further, the role fo nontrivial solutions to
Eq.~(\ref{eq:Ai-master-eq}) is not clear.
In this sense, the set up of the above ``zero-mode problem'' is
still ambiguous in general case.
To clarify this ``zero-mode problem'' itself, it will be
necessary to discuss this problem through the concrete
background metric $g_{ab}$ and appropriate boundary conditions
at $\partial\Sigma$ at first.
We will leave these issues as future works.


\section*{Acknowledgments}


The author deeply thanks to Prof. Masa-katsu Fujimoto for
various supports.




\end{document}